\documentclass[sigconf]{acmart}

\usepackage{booktabs} % For formal tables

\copyrightyear{2018} 
\acmYear{2018} 
\setcopyright{licensedothergov}
\acmConference[COMPASS '18]{ACM SIGCAS Conference on Computing and Sustainable Societies (COMPASS)}{June 20--22, 2018}{Menlo Park and San Jose, CA, USA}
\acmBooktitle{COMPASS '18: ACM SIGCAS Conference on Computing and Sustainable Societies (COMPASS), June 20--22, 2018, Menlo Park and San Jose, CA, USA}
\acmPrice{15.00}
\acmDOI{10.1145/3209811.3209879}
\acmISBN{978-1-4503-5816-3/18/06}

\begin{document}
\title[Exploiting Data and Human Knwl for Pred Wildlife Poaching ]{Exploiting Data and Human Knowledge for Predicting Wildlife Poaching}

\author{Swaminathan Gurumurthy}
\affiliation{%
  \institution{Carnegie Mellon University}
  \country{USA}}
\email{sgurumur@andrew.cmu.edu}
\author{Lantao Yu}
\affiliation{%
  \institution{Shanghai Jiao Tong University}
  \country{China}}
\email{yulantao@apex.sjtu.edu.cn}
\author{Chenyan Zhang}
\affiliation{%
  \institution{Southeast University}
  \country{China}}
\email{cy.zhang@seu.edu.cn}
\author{Yongchao Jin}
\affiliation{%
  \institution{World Wild Fund for Nature, China}
  \country{China}}
\email{jinyongchao@orientscape.com}
\author{Weiping Li}
\affiliation{%
  \institution{World Wild Fund for Nature, China}
  \country{China}}
\email{wpli@wwfchina.org}
 \author{Xiaodong Zhang}
\affiliation{%
  \institution{Huang Ni He Forest Bureau}
  \country{China}}
\email{1009595567@qq.com}
\author{Fei Fang}
\affiliation{%
  \institution{Carnegie Mellon University}
  \country{USA}}
\email{feifang@cmu.edu}

\renewcommand{\shortauthors}{S. Gurumurthy et al.}

\begin{abstract}
Poaching continues to be a significant threat to the conservation of wildlife and the associated ecosystem. Estimating and predicting where the poachers have committed or would commit crimes is essential to more effective allocation of patrolling resources. The real-world data in this domain is often sparse, noisy and incomplete, consisting of a small number of positive data (poaching signs), a large number of negative data with label uncertainty, and an even larger number of unlabeled data. Fortunately, domain experts such as rangers can provide complementary information about poaching activity patterns. However, this kind of human knowledge has rarely been used in previous approaches.

In this paper, we contribute new solutions to the predictive analysis of poaching patterns by exploiting both very limited data and human knowledge. We propose an approach to elicit quantitative information from domain experts through a questionnaire built upon a clustering-based division of the conservation area.
%derived from clustering the based on geographical features and asking domain experts to provide an estimate for each cluster. 
In addition, we propose algorithms that exploit qualitative and quantitative information provided by the domain experts to augment the dataset and improve learning. In collaboration with World Wild Fund for Nature, we show that incorporating human knowledge leads to better predictions in a conservation area in Northeastern China where the charismatic species is Siberian Tiger. The results show the importance of exploiting human knowledge when learning from limited data.
\end{abstract}

\begin{CCSXML}
<ccs2012>
<concept>
<concept_id>10003456.10003457.10003458.10010921</concept_id>
<concept_desc>Social and professional topics~Sustainability</concept_desc>
<concept_significance>500</concept_significance>
</concept>
<concept>
<concept_id>10010147.10010257.10010258.10010259.10010263</concept_id>
<concept_desc>Computing methodologies~Supervised learning by classification</concept_desc>
<concept_significance>300</concept_significance>
</concept>
<concept>
<concept_id>10010147.10010257.10010282.10011305</concept_id>
<concept_desc>Computing methodologies~Semi-supervised learning settings</concept_desc>
<concept_significance>300</concept_significance>
</concept>
</ccs2012>
\end{CCSXML}

\ccsdesc[500]{Social and professional topics~Sustainability}
\ccsdesc[300]{Computing methodologies~Supervised learning by classification}
\ccsdesc[300]{Computing methodologies~Semi-supervised learning settings}

\keywords{Computational Sustainability, Wildlife Conservation, Applied Machine Learning, Limited Labeled Data}

\maketitle

\section{Introduction}

Poaching is a profitable crime that threatens endangered species. Trade of ivory, tiger bones and skins are well-known examples of some of the incentives for this crime \cite{biggs2016elephant,saif2016poaching}. Wildlife conservation agencies aim to protect the wildlife and their habitats from poaching and illegal trade. To achieve this goal, they send rangers to patrol in protected conservation areas \cite{lemieux2014situational}. However, because of limited patrolling resources, it is impossible to monitor all intrusion routes and protect the entire area. Therefore, it is critical to understand how poachers make decisions by predicting the patterns of poaching to strategically allocate patrolling resources in order to detect and deter poaching activities and reduce potential harm to wildlife and their habitats.

Rangers record their findings, including animal signs and poaching activity signs, e.g., snares placed by poachers during the patrol, and therefore one can analyze these records to get insights of the poaching patterns.
There has been several previous work that provided predictive tools through designing machine learning models trained and evaluated using real-world data from two conservation sites in Uganda \cite{nguyen2016capture,kar2017cloudy,gholami2017taking,gholami2018adversary}.
In these work, a dataset is created based on past patrols and geographical information. The conservation area is discretized into a grid, and each data point corresponds to a grid cell during a particular time period, with the label being whether or not there were poaching activities in the cell in that time period. However, due to the limited patrolling resources and the inability of the rangers to detect all poaching activities signs during the patrols, the dataset often consists of a small number of positive data (cells with poaching activities found by rangers), a large number of negative data with label uncertainty (cells patrolled but with no poaching activities found), and an even larger number of unlabeled data (cell not patrolled).

Despite the effort made towards addressing these challenges, the sparsity of positive data is still a major challenge and the previous works failed to (i) exploit the unlabeled data to assist the learning; (ii) exploit human knowledge from domain experts, such as conservation site managers and rangers in a quantitative way.

The rangers have extensive experience in the field and have been interacting with poachers for years. They typically choose patrol routes with the goal of detecting and deterring poaching activities, and therefore the unlabeled data implicitly encodes their understanding about where the poaching activities may not take place. In addition, their knowledge about the poachers' capabilities, tactics, and the area goes beyond what is recorded in the dataset. The information they can provide is critical to building a better understanding of the poachers.
The key challenges here are to elicit quantitative information from them and to make efficient use of the information. 

In this paper, we focus on exploiting both the data and human knowledge to predict poaching activities. We provide the following contributions: (1) We present an approach to elicit quantitative information about poaching threat from domain experts.
We use k-means clustering to divide the conservation area into small regions or clusters and present the clusters to the domain experts. In the questionnaire, we ask them to provide an estimate of the poaching threat for each cluster. (2) We provide two approaches for exploiting human knowledge to enhance machine-learning-based predictive analysis. The first approach is to sample negative data points from unlabeled data. The second approach is to sample positive and/or negative data points from the cluster-based estimation of poaching threat. (3) We evaluate our proposed approaches using data from 2014-2018 in a conservation area in China. Our experimental results show that incorporating human knowledge can improve the performance of the predictive model. 

While we use anti-poaching as the domain of focus, the methodologies introduced in this paper are general and applicable to a large number of security problems with limited real-world data and experienced human experts, such as predicting crimes in areas with insufficient record.

\section{Related Work}
There has been some previous work focusing on understanding and predicting poaching activities. \cite{moore2017ranger} identifies that areas with certain features are at high risk of poaching and demonstrates that increasing the number of patrols will effectively reduce poaching activities using data from Nyungwe National Park. Similar work has been done at Tsavo National Park, \cite{shaffer2016predicting} performs geospatial analysis on the physical environment to find correlations between geographical features and poaching incidents. More recently, machine learning based approaches have been explored to predict poaching. Based on real-world data from Queen Elizabeth Protected Area, \cite{nguyen2016capture} uses a Dynamic Bayesian Network that explicitly models the dependencies between occurrence and detection of poaching activities, as well as the temporal pattern of poaching. \cite{kar2017cloudy} designs an ensemble of decision trees which incorporate spatial correlation of poaching to account for the undetected instances. \cite{gholami2017taking} provides a hybrid model that combines decision trees and Markov Random Fields \cite{gholami2017taking} to exploit the spatio-temporal correlation of poaching activities. \cite{gholami2018adversary} proposes to weigh the negative data points in the training set based on patrol effort so as to account for the label uncertainty. However, the challenge of having limited data is not fully resolved and human knowledge is only used in very limited ways in previous works such as to select features to be considered and to represent the dependency and correlation relationships.

In addition to wildlife poaching, predicting other types of crimes based on real-world data has been studied using general principles such as ``crime predicts crime'' in criminology. Machine learning models and algorithms including Bayesian networks, k-nearest neighbors, neural networks, decision trees, and support vector machines \cite{shojaee2013study,kang2017prediction} have been applied. However, most of these rely on a sufficiently large dataset and ignore the fact that there are undetected or unreported crime instances. In many domains including wildlife poaching, real-world data that can be used to build a machine learning model for predicting crime pattern is very limited and sometimes incomplete. Learning from such datasets and making a prediction is challenging.

Efforts have been made to study learning with limited labeled data and incorporating domain knowledge in the learning phase. For example, \cite{Nitesh} provides empirical studies to examine several aspects including independence or dependence amongst features, the amount of unlabeled data versus the amount of labeled data, and the effect of label noise. \cite{TingYu} proposes a guideline to incorporate domain knowledge into inductive machine learning while pointing out the difficulty in finding a universal solution, given the variety and diversity of domain knowledge. Progress has been made to address the problem. \cite{ZhipengLuo} introduces an approach to learn an instance-based classification model from subgroups of instances and a human estimation of the proportion of instances with certain labels. A method to learn from uncertain query responses from domain experts is also proposed by \cite{Taewan}. \cite{Aniket} gives a method to integrate domain knowledge in the field of breast cancer risk assessment by adding virtual instances to the original dataset. We leverage insights from this work to design an approach to elicit and exploit human knowledge.

\section{Domain Description and Real-World Dataset}
Wildlife poaching is a global challenge and poaching activities differ in many ways, including the poachers' goals, target species, tools, tactics and capabilities. For example, in 2012, poachers armed with grenades and AK-47s killed more than 300 elephants in Bouba Ndjidah National Park, Cameroon \cite{NGpoaching2014}. In 2012-2013, 70 lions died after eating poisoned food prepared by poachers in southern Africa \cite{VOA2017}. Among the different types of poaching activities, snare poaching is one of the most prevalent ways of poaching and is the main focus of this paper. Due to the low cost of making a snare (as low as \$0.5-\$2 per snare based on private conversations with experts from WWF) and the low skill requirement, snare poaching crimes committed by organized poachers as well as local villagers threaten various species in many countries across continents. 

Supported by government and non-government conservation agencies, rangers conduct patrols in order to detect snares before the wildlife gets trapped and deter the poachers. Rangers take records as they patrol, and in recent years, many conservation sites have started using software tools such as MIST\footnote{http://www.ecostats.com/web/MIST} and SMART\footnote{http://smartconservationtools.org/} to track and store the patrol routes and the records. However, the records only include poaching activities that are detected by the patrollers during their patrols. Some poaching activities were not recorded if the patroller did not patrol in the right area at the right time or did not find the well-hidden signs of poaching. Due to the limited patrolling resources and equipments, the records only cover a small portion of the poaching activities. In addition, wildlife animals are silent victims of poaching, making it hard to find other sources of reports on poaching activities. 

\subsection{Construction of Dataset}
In this paper, we focus on Huang Ni He National Nature Reserve (HNHR), a conservation area spanning about 75 sq. km in Northeastern China where the charismatic species is Siberian Tiger. Snare poaching is the main poaching activity of concern in this area. The rangers do not go on patrols all year around, but instead mostly patrol in autumn and winter seasons as there is a high poaching threat in these seasons. Based on geographical data, historical patrol records in 2013-2017, and recent records of 2017-2018 winter season patrol in HNHR, we divide the area into a 1km grid and construct a dataset where each data point corresponds to a grid cell in a patrol season. The label for each data point indicates whether or not there were any poaching activities in the grid cell in that patrol season. The features of each data point include: the distance from each area to the closest stream, village, patrol post, river, marsh, village road, provincial road, national road, highway, conservation boundary; land type, the elevation and slope of each area; patrol length (total distance patrolled in the grid) in last patrol season. Thus, we use a total of 14 features, of which 10 are distance features. 
%Patrolling and poaching data has been recorded using SMART starting from 2014. 
%Along with the amount of patrolling effort
% in each area, the datasets contain 14 years (2003-2016) of the
% type, location, and date of wildlife crime activities. To study wildlife
% crime, we divide the protected areas into 1 sq. km grid cells. Each
% of these cells is associated with several static geo-spatial features
% such as terrain (e.g., slope), distance values (e.g., distance to border,
% roads, and towns), and animal density. Additionally, each cell is
% associated with dynamic features such as patrol effort (coverage)
% across time and observed illegal activities (e.g., snares). Patrol effort
% is the amount of distance walked by park rangers across a cell at a
% specific time step. Since park rangers do not have unlimited manpower
% to patrol each cell thoroughly, it is possible that the amount
% of distance walked by them is not sufficient and consequently, some
% of the well-hidden snares are not detected by them. This fact is the
% source of uncertainty over the negative instances of crime and has
% to be considered in the adversarial reasoning

Here we provide details about how the dataset is constructed:\\
(i) \emph{Geographical Features}: Based on publicly available digital elevation model (DEM) data of the conservation area with 90m resolution \footnote{http://www.gscloud.cn/}, we derive the elevation and slope of the whole area through functions in Spatial Analyst Tools in ArcMap 10.2.
%aspect, \emph{aspect}
We get the land type and detailed locations of road, river and villages in the format of shapefiles by digitalizing the forest map of HNHR manually.
%in the format of shapefiles
%To get various distance features used in our model, we first use the MMQGIS plugin to create a grid layer and based on the grid layer, we use NNjoin plugin to extract the distance features from the corresponding shapefiles like rivers, roads and villages. 
To get various distance features used in our model, we create a grid using \emph{MMQGIS} tool in QGIS 2.14 and use \emph{NNjoin} tool to compute the distance from the nearest rivers, roads and villages.
To ease the training of the model, we further normalize the geographical feature values into the range of 0 to 1 except for land type feature.\\
%\paragraph{Protect sensitive data}
% Not sure what to write in this part
(ii) \emph{Patrol Features}:
Through the anti-poaching project supported by WWF, historical patrol routes as well as animal activities and poaching activities detected during patrols are recorded in SMART. Based on the exported file of historical waypoints, we compute the patrol length in each grid cell by splitting the patrol line into fixed intervals using \emph{QChainage} tool and counting the intervals using Analysis Tools (\emph{Count Points in Polygon} function) in QGIS 2.14. We also normalize the patrol length feature into the range of 0 to 1.\\
%split polylines into fixed 1500 m intervals
%extract the historical patrol length in each grid cell using the QChainage plugin and Analysis Tools in QGIS.
%``count points in polygon'' function
%QChainage is either separating the selected (or all) Line-Features into parts, or linear referencing after a chosen distance. All along the whole Feature or between selectable start and endpoint.
%To get the animal density and poaching activities information, we import the obervation data into QGIS and use the ``join attributes by location'' function.
(iii) \emph{Labels}: To get the label of each data point, we count the detected poaching activities by importing the SMART observation data into QGIS and performing spatial joins using Data Management Tools (\emph{Join Attributes by Location} function) in QGIS 2.14. The labels are then binarized by setting all areas with non-zero poaching activity detections to one (Positive Data Points) and the rest to zero (Negative Data Points).

% \subsection{Deployment}
% Since the dataset is very small, we recognize that the validation performance might not be a very accurate representation of real world performance. Hence, we deployed an initial version of our model, Bagging Ensemble decision trees with Negative Sampling, in a real life conservation area for Tiger protection. The patrol effort went on for 3 months based on our threat predictions. The patrollers found 7 threats over 34 rounds of patrolling with an average of 0.2 threats per patrol effort. Each patrol effort took 10265.24 seconds (~2.85 hours) on average. 
In addition to historical data collected from 2013-2017, we also get recent records in 2017-2018 winter season in HNHR, which are collected through 34 patrols that are partially informed by predictions made by an initial version of the trained model (see Section \ref{sec:eval} for more details).
%which are collected through 34 patrol shifts spanning over 3 months of patrol. Each patrol shift took about $2.85$ hours on average. 

\subsection{Challenges in the Dataset}
The lack of patrolling resources leads to a dataset with several challenges and peculiarities. It suffers from significant class imbalance, sparsity and noise in negative labels.

\begin{figure}[t]
\caption{Number of Positive Examples}
\centering
\includegraphics[width=0.4\textwidth]{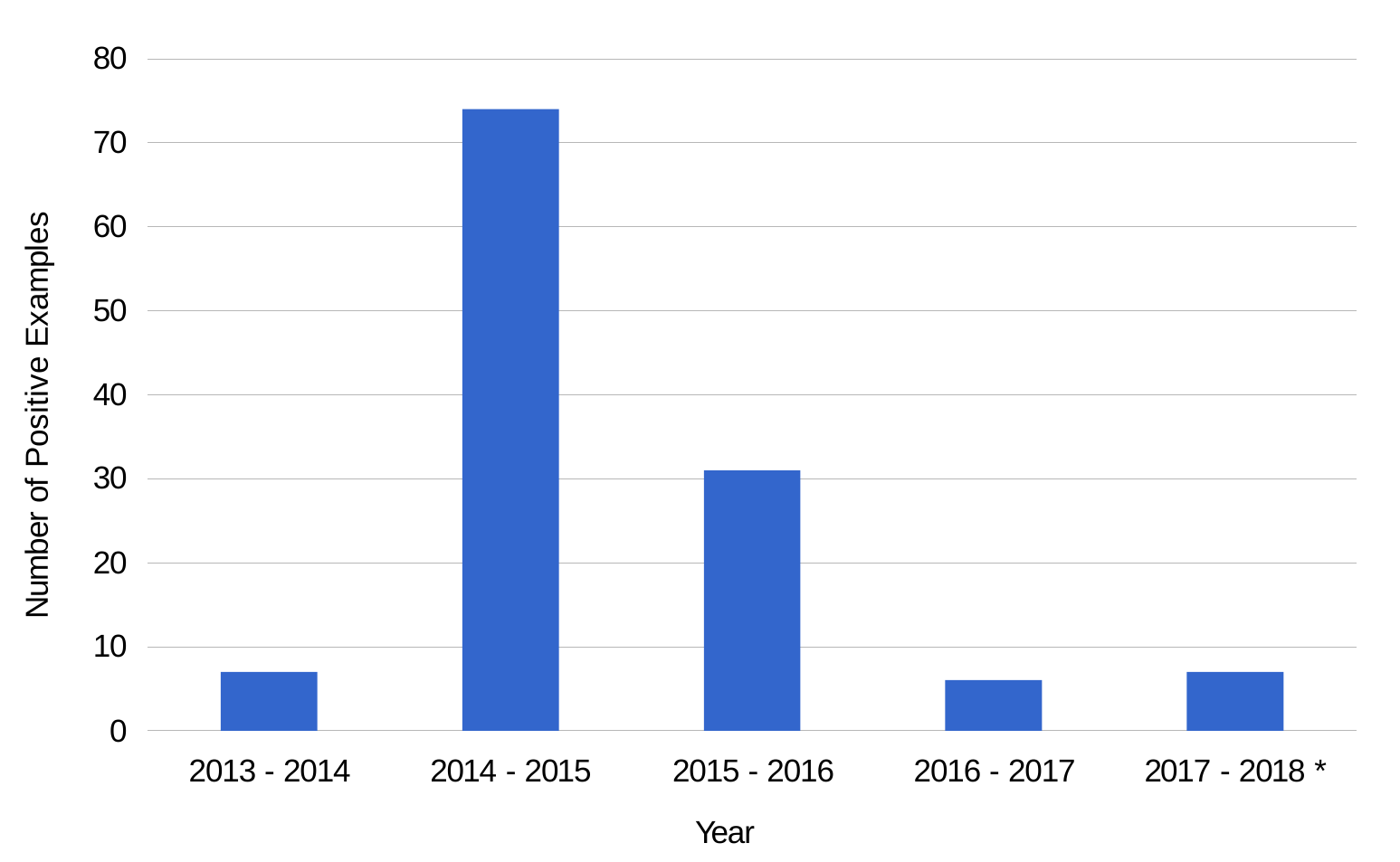}\label{fig:Positive}
\end{figure}

\begin{figure}[t]
\caption{Number of Negative Examples}
\centering
\includegraphics[width=0.4\textwidth]{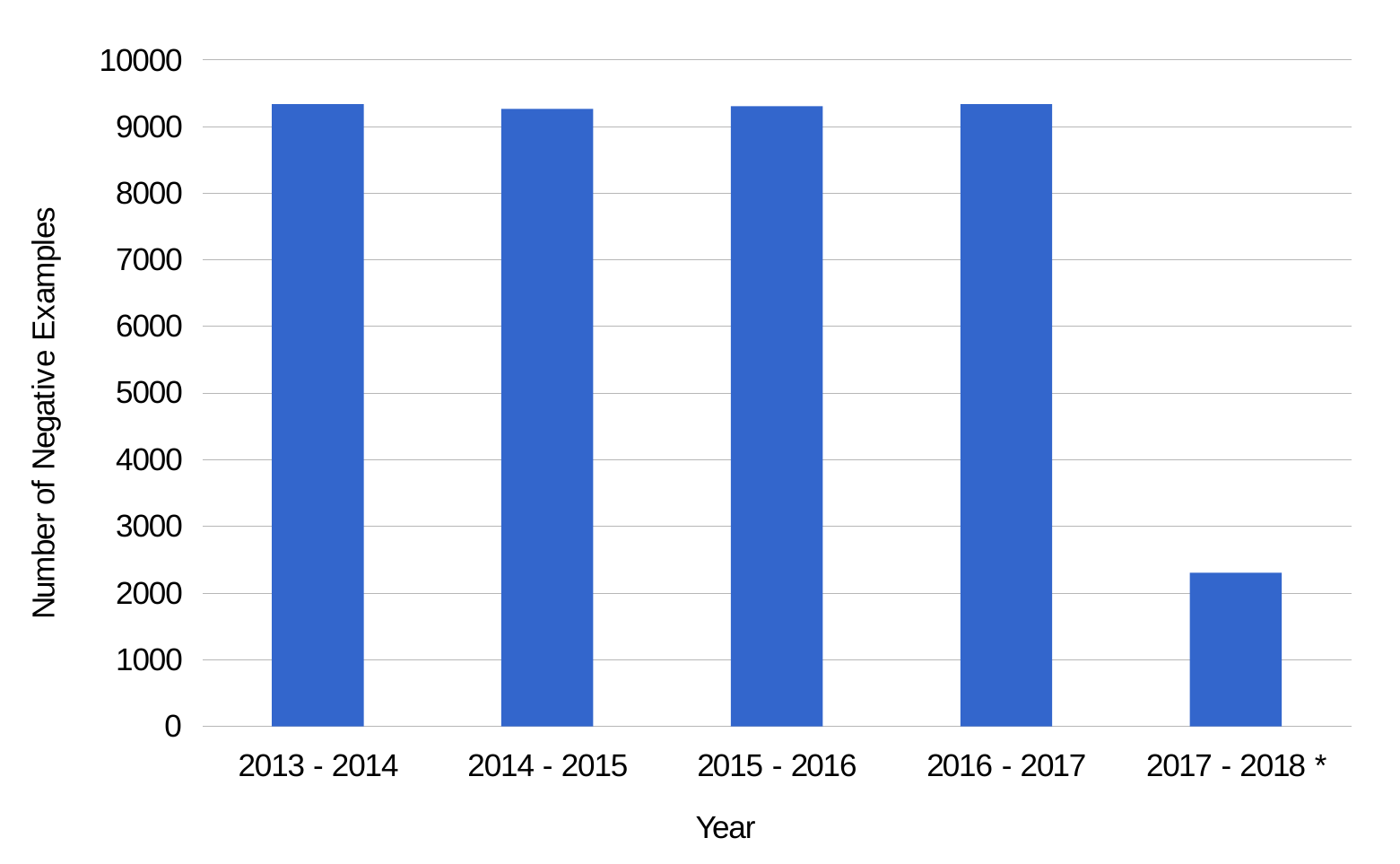}\label{fig:Negative}
\end{figure}

\begin{figure}[t]
\caption{Number of Unlabeled Examples}
\centering
\includegraphics[width=0.4\textwidth]{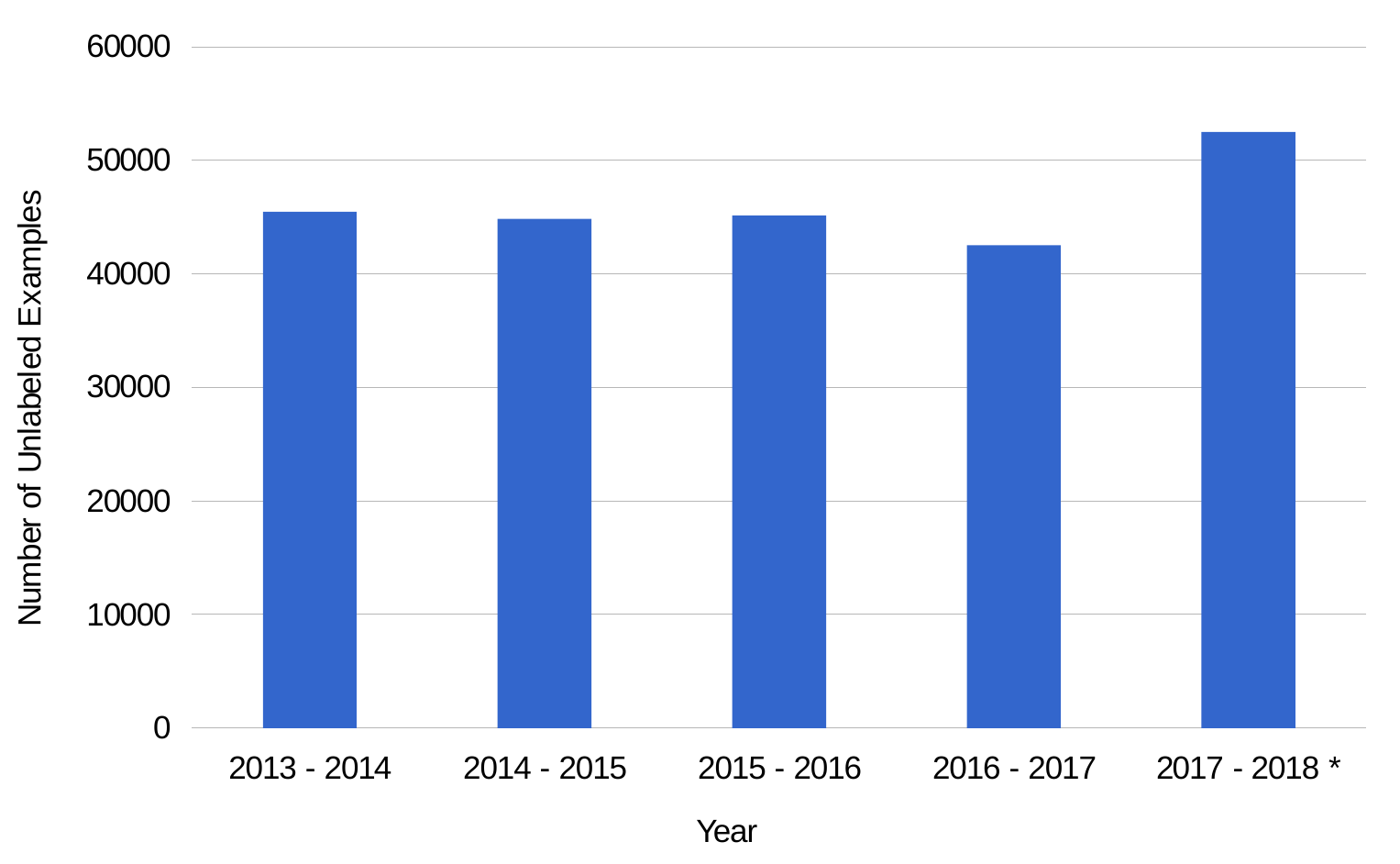}\label{fig:Unlabeled}
\end{figure}

In fact, as can be seen in Figure \ref{fig:Positive}-\ref{fig:Unlabeled}, the dataset is extremely skewed with very few positive examples ($29.5$ on average\footnote{The data in 2017-2018 is not complete and is excluded from the computation of average.}) in contrast to a large number of negative examples ($9310.5$ on average). In addition, only a small portion of the area is patrolled every year, leading to a very larger set of unlabeled data ($44493.25$ on average). 

\begin{figure}[t]
\caption{A snare that is hard to detect.}
\centering
\includegraphics[width=0.3\textwidth]{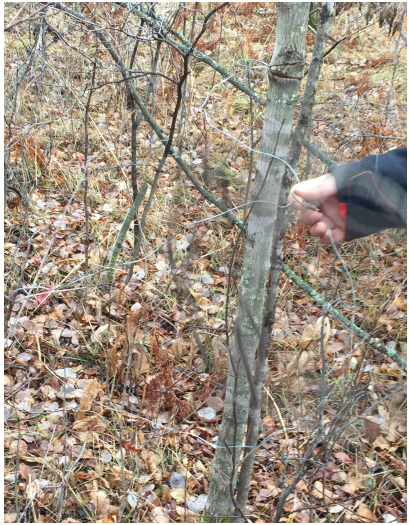}\label{fig:snare}
\end{figure}

Detecting snares requires experience and in many cases, the snares cannot be found easily, given the tree branches in the scene, especially when one is walking. Figure \ref{fig:snare} shows an example snare that cannot be easily detected by an unexperienced ranger. As a result, the patrollers might have simply missed the snares in certain regions while patrolling and hence a lot of the negative data points might indeed be positive. Therefore, we expect noise in the negative labels in our dataset.
 
%For example, less than 150 poaching activities are recorded in a conservation area in Northeastern China in 2016 (based on conversation with our partners from WWF) but there are undetected poaching activities due to the limited patrol resources and silent victim issue. 


%Talk about the size, problems with evaluation, noisy labels, imbalanced dataset.

\section{Methodology}
Given the dataset, we aim to train a model that can predict or estimate the label for any data point using geographical and patrol features.
%We seek to model this problem as a classification problem, where we use the features calculated to estimate the label for each data point. 
To help address the challenges of the dataset, we propose several approaches to elicit and exploit human knowledge. First, we collect quantitative domain knowledge from experts through questionnaires built upon clustering. Second, we use the collected quantitative domain knowledge to augment the dataset. Third, we use data duplication to alleviate the dataset skew towards negative examples. Fourth, noting that most of the unlabeled data are negative data, we augment the negative dataset with a randomly sampled subset of the unlabeled data. We will discuss these approaches in more detail in subsequent sections. 

Several different machine learning models have been proposed previously to predict poaching threats. However, this paper focuses on using expert knowledge to enhance machine learning based predictive tools. Therefore, we apply our approaches to two basic machine learning models to illustrate the importance of using expert knowledge: a bagging ensemble of decision trees and a neural network based model. %With the results using these two exemplary models, we expect improved performance of more complex models when our approach is applied.

%Empirically, the model that performs best in our experiments was a bagging ensemble of decision trees. For comparisons, we also implement other baseline models like Neural Networks and analyze their performance on this dataset.

%Empirically, the model that performs best in our experiments was a bagging ensemble of decision trees. For comparisons, we also implement other baseline models like Neural Networks and analyze their performance on this dataset.

\subsection{Eliciting Information from Domain Experts}
\label{sec:elicitinfo}
Given the peculiarities of the dataset, it is very important to elicit some explicit or implicit information from domain experts so that we can make use of the unlabeled data as well as the noisy labeled data. It should be noted that several factors need to be kept in mind while collecting domain knowledge from the experts. Firstly, it is not possible for the experts to give very accurate and fine-grained information (e.g, the specific probabilities for every region). Second, the experts cannot be expected to label a huge amount of data given the limited amount of resources and time they have. Third, it should be expected that the information provided by the experts can be noisy. Hence, we have to settle for a limited amount of information which is coarse-grained and noisy. But at the same time, we need to ensure that we are able to extract enough information to tackle the bottlenecks in the machine learning models. With these in mind, we develop the following method to elicit information from the  domain experts.
%additionally talk about other possible approaches considered and not chosen? like 

\begin{figure}[t]
\caption{Visualization of the 40 clusters}
\centering
\includegraphics[width=0.5\textwidth]{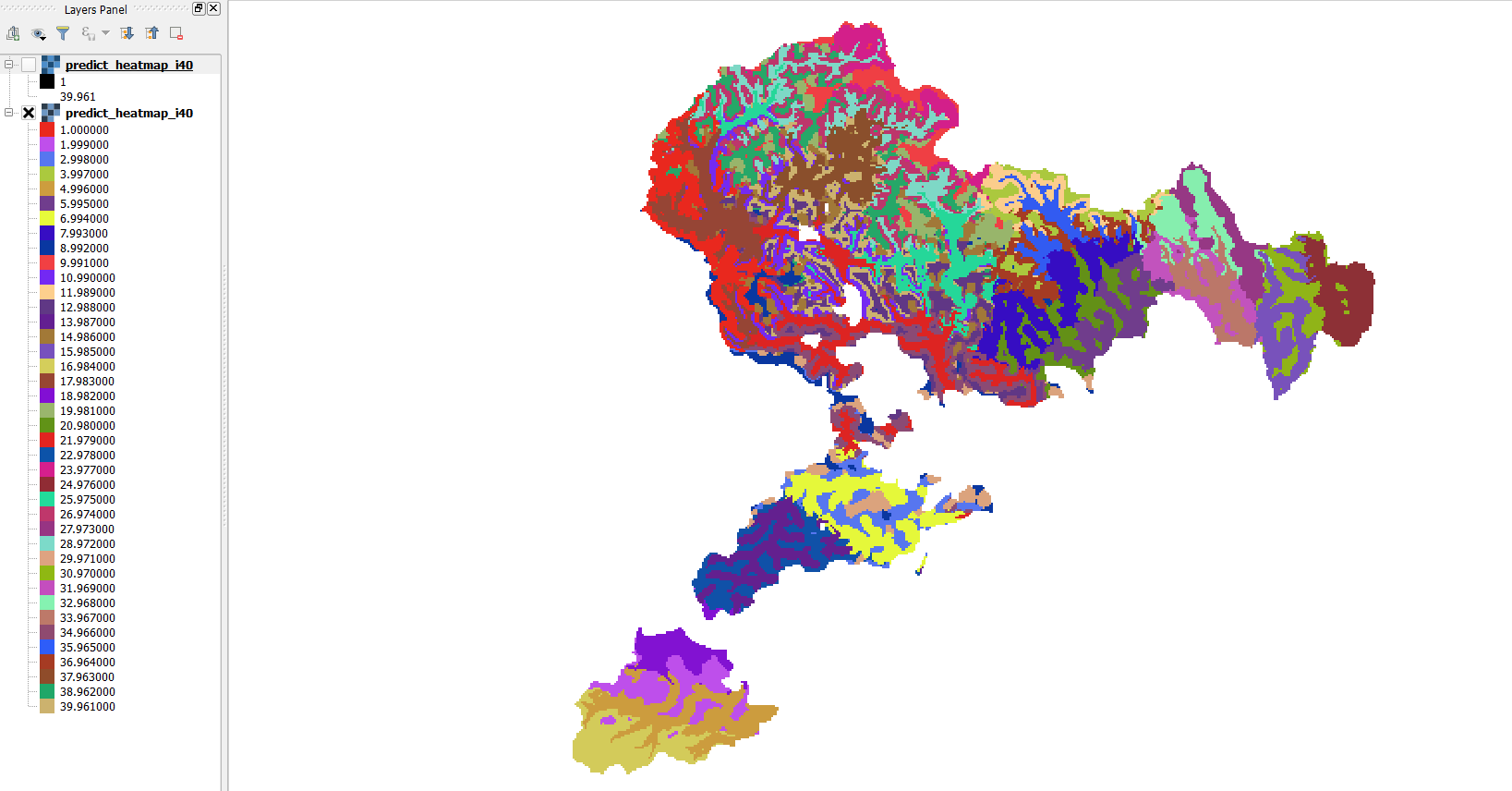}\label{fig:40cluster}
\end{figure}

%\paragraph{Cluster Labels}
Recognizing the fact that it is difficult for the experts to provide estimates of poaching threat level for every single grid cell in the conservation area, and that fine-grained labels provided by the experts can be very noisy, we propose a method to obtain information at a coarse level from the experts. Instead of asking the experts to score each individual cell, we first group these cells into clusters using K-means clustering in the feature space. We then present these clusters to the experts and ask them to provide a score for each cluster from $1$ to $10$, where $1$ corresponds to minimum threat level and $10$ corresponds to maximum threat level. To decide the number of clusters to be used in this procedure, we asked the domain experts for feedback. They believed that $40-50$ clusters would be a reasonable number to ensure consistency across the labeling on a given set of clusters.
Following their advice, we repeat this procedure twice, first with $40$ clusters and then with $50$ clusters. Doing it twice helps account for any inconsistency in the scores provided by the experts. This gives us two sets of clusters and their corresponding scores. This approach helps us extract useful information about the threat level without causing much cognitive burden on the experts. Specifically, now the experts only need to provide $90$ scores rather than about $75000$ scores, one for each grid cell. 
See Figure \ref{fig:40cluster} for visualizations of the 40 clusters in the map. 

\subsection{Constructing Aggregated Score}
\label{sec:score}
Based on the two set of scores provided by the expert, we compute the aggregated score as an indicator of the threat level of a grid cell.
%\paragraph{Cluster Labels}
The experts provide us with a score (from $1$ to $10$) for each cluster for each questionnaire. But we have these confidence score for each cluster set individually. In order to combine these two sets of cluster scores, we propose a simple approach. If a grid cell $k$ belongs to a cluster $C_i^1$ with score $s_1(C_i^1)$ when 40 clusters are used and belongs to a cluster $C_j^2$ with score $s_2(C_j^1)$ when 50 clusters are used, we define the aggregated score as the minimum of the two, i.e., 
\begin{equation*}
    s(k) = \min\{s_1(C_i^1),s_2(C_j^1)\}
\end{equation*}
%where $min(a,b)$ gives the minimum of a and b. 
In other words, we assign a high score to a data point only if it received a high score in both cluster sets.

\subsection{Data Augmentation}
As discussed earlier, the dataset has several unique properties that make the straightforward application of a machine learning algorithm lead to poor performance (see Section \ref{sec:eval}). These issues can be mitigated to an extent using data augmentation. In this work, we propose three ways for augmenting the data.

\paragraph{Data Duplication (DD)}
Since the number of positive examples are very low compared to the negative examples, we choose to duplicate the positive examples to balance the dataset. With this approach, we ensure that the number of positive examples and negative examples used for training the model remains the same level.

\paragraph{Negative Sampling (NS)}
As established previously we know that most of the unlabeled examples are low threat regions, since the experts chose not to explore those regions. Hence we add a random partition of the unlabeled set into the negative set during training. It is important to note here that unlabeled data points have never been patrolled and are thus year agnostic. Thus we only add the one data point per grid when sampling unlabeled data points instead of adding a sample for each year. 
%\paragraph{Dynamic Negative Sampling}
%As discussed in the previous section it might be useful to treat some portion of unlabeled set as negative data. But instead of doing random sampling, it has been shown that in various settings it's more useful to iteratively sample points from the unlabeled set which the classifier thinks are positive examples. This helps especially if the classifier things a significant number of the unlabeled samples are positive and helps remove this bias more effectively.
\paragraph{Score based Positive Sampling (PS)}
Although the information collected from domain experts as described in Section \ref{sec:elicitinfo} is coarse and noisy, it can be incorporated into the machine learning model in a variety of ways. In this work, we propose to use the aggregated score and add the unlabeled data points that are likely to have positive labels to the positive dataset. 
We add all unlabeled data points whose corresponding grid cell has aggregated scores greater than or equal to 6 to the positive dataset. Similar to NS, we add only one sample per grid when sampling from unlabeled data points instead of adding an entry for each year.

The information collected from domain experts can also be used to estimate the probability of each unlabeled data point being positive, or to build a weighing scheme over the data points to be used in the loss function when training the model, or to be added as a regularization term in the objective function when training the model. We leave these options for future investigation. 

%\paragraph{Prior Knowledge for Negative Sampling}
%Similar to sampling positive data points using the score obtained from the prior, we can also sample negative data points using the scores and sample the most negative data points. This strategy can sometimes induces a bias from the prior knowledge. Also, if the prior is good, then this might not help because the samples added won't be hard negatives

\subsection{Model Implementations}
\paragraph{Bagging Ensemble Decision Tree}
We use bagging ensemble decision tree \cite{breiman1996bagging} with 1000 trees where each base tree is trained using only 10 percent of the total training data. We use entropy to compute the information gain at each node. We use this as our default classifier due to the superior performance (empirically demonstrated in the experiments). We use the implementation provided scikit-learn with the above mentioned parameters to train the model. 
\paragraph{Neural Networks}
We also use a three-layer feedforward neural network \cite{lecun2015deep} with 8 neurons on the first layer , 4 neurons on the second layers and a single neuron in the last layer spitting out the threat probability for that data point. We use relu nonlinearity in the first and second layers and a sigmoid at the output. To predict the final output we use an ensemble of 100 such neural networks. We train using the Adam optimizer with learning rate $= 0.001$, beta1$=0.9$ and beta2$=0.999$. We keep a constant batch size of 64 throughout training. We also add an l2 penalty on the weights with weight decay constant of 0.1. 

\section{Evaluation}
\label{sec:eval}
In this section, we evaluate our model and showcase the importance of each component of the model. We compare these models on 4 commonly used metrics. 
An early version of the model which uses decision tree ensemble with DD and NS has been used to inform patrols in a pilot field test in October 2017, and in a subsequent field test spanning over 3 months of patrol from November 2017 to February 2018. 
%We also present the field test results. 
The data collected from these field tests are also merged into the dataset while we report the results. A more detailed discussion about the field tests can be found in Section \ref{sec:fieldtest}. Evaluation of the models on the data collected during 2013-2017 can be found in Section 7.

\subsection{Metrics}
We evaluate our model performance using 4 metrics precision, recall, F1 score and the ll score. We choose to report the ll score along with the precision, recall and F1 scores because it offers better discriminability since it's not bounded between $[0,1]$. We do not include the more commonly used AUC ROC score because we observed that it was sensitive to false positives and gave a high score even when the number of false positives were high. 
\begin{equation*}
  Precision = \frac{True Positives}{True Positives + False Positives}
\end{equation*}
\begin{equation*}
  Recall = \frac{True Positives}{True Positives + False Negatives}
\end{equation*}
\begin{equation*}
  F1 score = \frac{2*Recall * Precision}{Recall + Precision}
\end{equation*}
\begin{equation*}
  ll score = \frac{Recall * TestSetSize}{True Positives + False Positives}
\end{equation*}

\subsection{Training and Testing}
Given the limited number of positive samples in the dataset, it is infeasible to separate the dataset into a test set and a training set without getting a biased evaluation of our model. Hence, we used 4-fold cross validation to train and test our model performance. But we still observed very high variance over the cross validation runs. Thus, we repeat 4-fold cross validation multiple times and average the results across all the runs to get a more accurate estimate of the scores. The dataset here includes the historical data collected by patrollers and the data collected during the field test of an earlier version of our model (with only decision trees and negative sampling). Cross-validation is performed on this merged dataset and the mean of precision, recall, ll score and F1 score are reported on each model after multiple evaluations.
\subsection{Evaluation on Dataset}
In this section we evaluate our model on the dataset and also show the impact of each component of the model. Table \ref{tab:ps1scores} contains the precision, recall, F1 and the ll scores for the model and multiple baselines. In the experiments listed in the Table \ref{tab:ps1scores}, DD indicates Data Duplication (i.e, duplicate the positive data to match the number of negative examples). We find that data duplication is very crucial. The model completely fails (predicts negative labels for every example) if we remove this component. We test a more sophisticated data oversampling technique called SMOTE: Synthetic Minority Over-sampling Technique \cite{SMOTE} to compare against data duplication. In this technique, new examples are created by interpolating between random pairs of k-nearest neighbor positive samples. We observe that this does not help with our dataset. We observe that the Positive Sampling (PS) when added standalone significantly deteriorates the ll score, F1 score and precision since it leads to an increase in the false positive rate. Adding Negative sampling (NS) standalone does not cause much benefit either. But adding both positive and negative sampling together leads to a boost in the precision, F1 and the ll score.  We also observe that the neural-network-based models performs poorly compared to the decision tree based ensemble. But the Positive Sampling (PS) results in performance improvement in the neural network as well. This shows that expert knowledge can help boost the performance of both the machine learning models even if their relative performances are very different. We observe that the neural network has a very high false positive rate even after training with Negative sampling. We also include the scores computed when using a random classifier which labels any example as positive with probability $0.5$ to give the readers a sense of baseline values for each of the scores.
\begin{table}
\caption {Scores for our model and contribution of each component} \label{tab:ps1scores}
\begin{center}
\begin{tabular}{ |c|c|c|c|c|} 
 \hline
Models & LL score & Recall & Precision & F1 score\\
\hline
Random decisions & 0.51 & 0.5 & 0.004 & 0.008 \\
DT & 0.0 & nan & 0.0 & 0.000 \\
DT with DD & 14.60 & 0.31 & 0.17 & 0.219\\ 
DT with SMOTE & 11.19 & 0.35 & 0.12& 0.179\\
DT with DD, PS & 4.99 & 0.35 & 0.05 & 0.087\\   
DT with DD and NS & 14.05 & 0.27 & 0.19 & 0.223\\ 
DT with DD, NS, PS & 15.42 & 0.31 & 0.18 & 0.227\\
NN & 0.0 & nan & 0.0 & 0.000 \\
NN with DD & 3.26 & 0.72 & 0.016 & 0.031 \\
NN with DD, NS & 2.47 & 0.48 & 0.02 & 0.038\\
NN with DD, NS, PS &  3.70 & 0.79 & 0.02 &  0.039 \\
\hline
\end{tabular}
\end{center}
\end{table}
\subsection{Alternative Domain Knowledge}
\begin{figure}[t]
\caption{Histograms for dist-village and dist-river}
\centering
\includegraphics[width=0.23\textwidth]{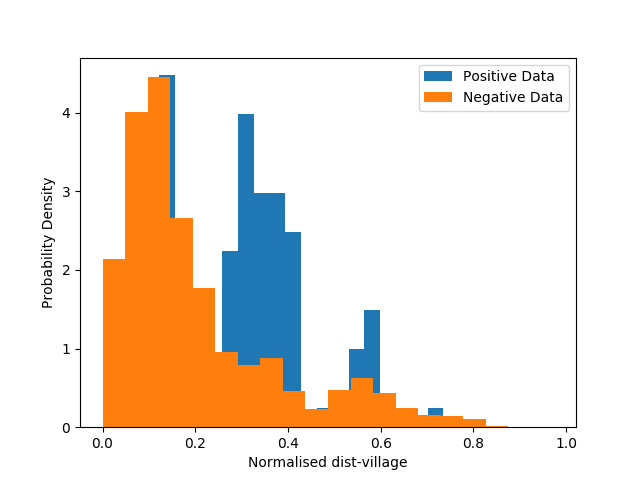}\label{fig:distv}
\includegraphics[width=0.23\textwidth]{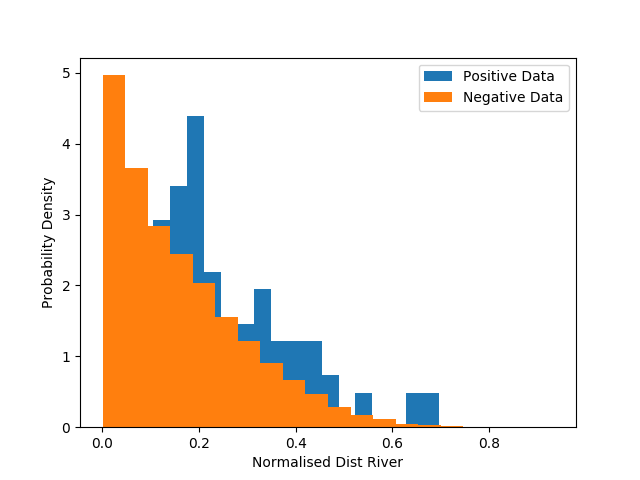}\label{fig:distr}
\end{figure}

We observed in the previous section that the model performance is very sensitive to the positive sampling strategy adopted. In fact to illustrate this we collect another type of data from the domain experts to estimate the confidence scores on unlabeled data. We plotted the histograms for individual features and found that in some of them there was a clear difference between positive and negative samples. The histograms for dist-village and dist-river can be seen in Figure \ref{fig:distv}. Hence, we asked the experts to provide specific ranges for some of the features which they think have high correlation with the labels. We also asked for their confidence on the specific ranges provided (in line with the 4-point estimates \cite{4point} to elicit distributions). However, the experts commented that providing confidence for each feature was difficult since multiple other factors need to be taken into account. Thus they provided only the ranges for 11 features. A sample of 3 of those features has been shown in Table \ref{tab:ftrranges} for illustration purposes. 

However, we observed that the ranges for some of these features do not correlate well with data. For example, the range given for dist-river does not match with the histogram shown in the plot. Hence, we also tested the case where we prune the features whose ranges do not correlate well with data. After the pruning, we were left with ranges for 4 features namely, dist-patrol, dist-village, elevation and slope. We tested the effectiveness of this approach for positive sampling to examine the utility of this type of expert knowledge. 

\begin{table}
\caption {Feature Ranges provided by Domain experts} \label{tab:ftrranges}
\begin{center}
\begin{tabular}{ |c|c|c|c|c|c|c|c|c|c|c|c| } 
 \hline
Snaring Threat & dist-village & dist-patrol & dist-river \\
\hline
low & [0,0.3] & [0,0.03] & [0.1,1] \\ 
high & [0.3,1] & [0.03,1] & [0,0.1] \\
\hline
\end{tabular}
\end{center}
\end{table}
%See Figure \ref{fig:40cluster} and Figure \ref{fig:50cluster} for visualizations of the clusters in the map. 

Given the feature ranges for which the probability $P(Y=1|X)$ is high, we aim to unlabeled data points which are more likely to have positive labels. We assume that, given feature $X_{i}$ and corresponding feature range, $F_{i}$, the probability, $P(Y=1|X_{i} \in{F_{i}}) = p$ and $P(Y=1|X_{i} \notin{F_{i}}) = q$, for any the features. We can apply Bayes rule, and use the Naive Bayes assumption on the features. Simplifying it we get, 
\begin{equation}
P(Y=1 | X^{(j)}) = \frac{\prod_{i=1}^{n} {P(Y=1|X_{i}^{(j)})}}{P(Y=1)^{n-1}} \\
\end{equation}
where $n$ is the total number of features we consider.
The denominator can be replaced with a normalization constant $Z$. If $m$ out of the $n$ features fall inside the specified ranges provided, the equation above can be simplified to
\begin{equation}
P(Y=1 | X^{(j)}) = \frac{p^{m}q^{n-m}}{Z} \label{eqn:prob}\\
\end{equation}
To estimate the probability of being positive for each unlabeled data point, we use $p=0.04$ and $q=0.01$ as advised by the experts. We pick 1000 samples that have the highest probability computed through Equation \ref{eqn:prob} and add them to the Positive dataset. We refer to this data augmentation approach as Positive sampling using feature ranges (PSFR) and report the scores in Table \ref{tab:ps2scores}. We observe that using these features brings down the performance of the model even after feature pruning. This shows how sensitive the performance is to positive sampling. Hence positive sampling in such cases needs to be done with utmost care.
\begin{table}
\caption {Scores when positive sampling using feature ranges} \label{tab:ps2scores}
\begin{center}
\begin{tabular}{ |c|c|c|c|c|} 
 \hline
Models & LL score & Recall & Precision & F1\\
\hline
DT with DD & 14.60 & 0.31 & 0.17 & 0.219 \\ 
DT with DD, PSFR & 6.87 & 0.34 & 0.07 & 0.116\\   
\hline
DT with DD and NS & 14.05 & 0.27 & 0.19 & 0.223\\ 
DT with DD, NS, PSFR & 10.88 & 0.29 & 0.14 & 0.189\\ 
\hline
\end{tabular}
\end{center}
\end{table}
% \subsection{Deployment}
% Since the dataset is very small, we recognize that the validation performance might not be a very accurate representation of real world performance. Hence, we deployed an initial version of our model, Bagging Ensemble decision trees with Negative Sampling, in a real life conservation area for Tiger protection. The patrol effort went on for 3 months based on our threat predictions. The patrollers found 7 threats over 34 rounds of patrolling with an average of 0.2 threats per patrol effort. Each patrol effort took 10265.24 seconds (~2.85 hours) on average. 

\subsection{Field Tests}
\label{sec:fieldtest}
The predictions of poaching activities made based on \emph{DT with DD and NS} trained on 2013-2017 dataset has been used to guide two sets of field tests.

\begin{figure}[t]
\caption{22 snares found during the pilot field test in October 2017.}
\centering
\includegraphics[width=0.3\textwidth]{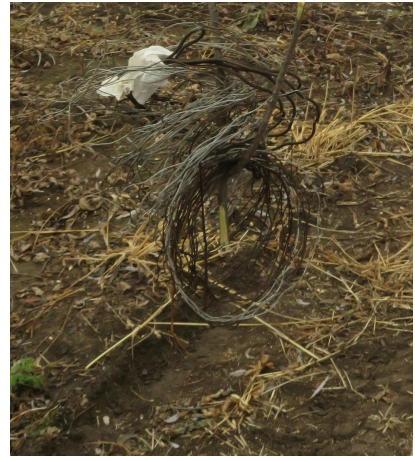}\label{fig:fieldtest}
\end{figure}

In October 2017, a two-day field test was conducted in HNHR.  The rangers selected two patrol routes that covered areas which had not been frequently patrolled but were predicted to have poaching activities by our model. During the field test, 22 snares were found (see Figure \ref{fig:fieldtest}), demonstrating the potential of a machine learning model enhanced with human knowledge. 

During November 2017 to February 2018, another set of field tests are conducted, consisting of 34 patrol shifts. Each patrol shift took about $2.85$ hours on average. During these patrols, 7 snares were found. However, rangers mentioned that that the low number of findings during these patrols is mainly due to the reduced tolerance to poaching in China this year, as can be seen from a set of changes in policy \cite{ChinaIvory}.

As can be seen from Table \ref{tab:ps1scores}, \emph{DT with DD, ND, PS} outperforms the version which does not use the expert based positive sampling but was used in these field tests.
We expect to run larger scale tests with this best-performing model that incorporates human knowledge elicited through the clustering-based questionnaires and data from conservation sites in Northeastern China in the next patrol season.

%We expect t

\section{Conclusion and Discussion}
%We observe that the data collected in conservation areas has several peculiarities. The data needs to be pre-processed properly to extract meaningful features that can be very useful for classification. The dataset even after pre-processing is very challenging. 
In this paper, we focus on eliciting and exploiting human knowledge to enhance the predictive analysis in wildlife poaching. The dataset in this domain has very few positive data points and suffers from uncertainty in negative labels.
We designed questionnaires to elicit information from domain experts. The information is then used to estimate a threat level of each data point. Based on the estimated threat level and other qualitative domain knowledge such as most unlabeled data points are negative, we augment the dataset for training. We apply our approach to a bagging ensemble of decision trees, which leads to significant improvement using multiple evaluation criteria. Improvements can also be seen when the develop approach that combines data and human knowledge is applied to a neural network model. However, decision tree ensemble lead to much better performance than the neural network based model. Taking cues from the results obtained on this exemplar model, we expect improved performance on more complex models as well, when our approach is applied.

Through our experiments, we show that making use of the human knowledge properly can lead to performance improvement in prediction. We find that the performance is very sensitive to positive data augmentation. Hence incorporating the expert knowledge to augment the positive data can be very useful. But we also observed that this needs to be done carefully. Biases in the expert supervision can affect the performance adversely in certain cases as shown in the Alternative Domain Knowledge section. We demonstrate a more principled method of collecting this prior knowledge by clustering the data points and getting weak labels over the clusters. But this can also become the bottleneck of the proposed approach. We observed that effective communication with the experts is crucial when relying on domain knowledge. But geographical/language barriers between the experts and the researchers can come in the way, making communication very difficult.

It is important to note that our approach is fairly generic and can be used in a variety of other settings, where the expert knowledge is costly and a large portion of the data is unlabeled. In future work, we hope to test our techniques in similar domains like wildlife monitoring and show the effectiveness of our approach. Another future direction is to explore other alternatives of collecting quantitative information from domain experts which can be effectively incorporated into machine learning models. 

\begin{acks}

We thank rangers from Huang Ni He Forest Bureau, China and officers from WWF China Northeastern Office. We thank Tianyu Gu and Justin Jia for their help in analyzing the data and for providing valuable feedback.

\end{acks}

\bibliographystyle{ACM-Reference-Format}
\bibliography{sample-bibliography}


\begin{thebibliography}{23}

%%% the Web address.

\ifx \showCODEN    \undefined \def \showCODEN     #1{\unskip}     \fi
\ifx \showDOI      \undefined \def \showDOI       #1{#1}\fi
\ifx \showISBNx    \undefined \def \showISBNx     #1{\unskip}     \fi
\ifx \showISBNxiii \undefined \def \showISBNxiii  #1{\unskip}     \fi
\ifx \showISSN     \undefined \def \showISSN      #1{\unskip}     \fi
\ifx \showLCCN     \undefined \def \showLCCN      #1{\unskip}     \fi
\ifx \shownote     \undefined \def \shownote      #1{#1}          \fi
\ifx \showarticletitle \undefined \def \showarticletitle #1{#1}   \fi
\ifx \showURL      \undefined \def \showURL       {\relax}        \fi
% invisible to TeX
\providecommand\bibfield[2]{#2}
\providecommand\bibinfo[2]{#2}
\providecommand\natexlab[1]{#1}
\providecommand\showeprint[2][]{arXiv:#2}

\bibitem[\protect\citeauthoryear{Aniket~Bochare}{Aniket~Bochare}{2014}]%
        {Aniket}
\bibfield{author}{\bibinfo{person}{Yelena Yesha Anupam Joshi Yaacov Yesha Mary
  Brady Michael A. Grasso Napthali~Rishe Aniket~Bochare, Aryya~Gangopadhyay}.}
  \bibinfo{year}{2014}\natexlab{}.
\newblock \showarticletitle{Integrating domain knowledge in supervised machine
  learning to assess the risk of breast cancer}.
\newblock \bibinfo{journal}{\emph{Int. J. Medical Engineering and Informatics}}
  (\bibinfo{year}{2014}).
\newblock


\bibitem[\protect\citeauthoryear{Bale}{Bale}{2017}]%
        {ChinaIvory}
\bibfield{author}{\bibinfo{person}{Rachael Bale}.}
  \bibinfo{year}{2017}\natexlab{}.
\newblock \bibinfo{title}{China Shuts Down Its Legal Ivory Trade}.
\newblock
  \bibinfo{howpublished}{\url{https://news.nationalgeographic.com/2017/12/wildlife-watch-china-ivory-ban-goes-into-effect/}}.
\newblock


\bibitem[\protect\citeauthoryear{Biggs}{Biggs}{2016}]%
        {biggs2016elephant}
\bibfield{author}{\bibinfo{person}{Duan Biggs}.}
  \bibinfo{year}{2016}\natexlab{}.
\newblock \showarticletitle{Elephant poaching: Track the impact of Kenya's
  ivory burn}.
\newblock \bibinfo{journal}{\emph{Nature}} \bibinfo{volume}{534},
  \bibinfo{number}{7606} (\bibinfo{year}{2016}), \bibinfo{pages}{179}.
\newblock


\bibitem[\protect\citeauthoryear{Breiman}{Breiman}{1996}]%
        {breiman1996bagging}
\bibfield{author}{\bibinfo{person}{Leo Breiman}.}
  \bibinfo{year}{1996}\natexlab{}.
\newblock \showarticletitle{Bagging predictors}.
\newblock \bibinfo{journal}{\emph{Machine learning}} \bibinfo{volume}{24},
  \bibinfo{number}{2} (\bibinfo{year}{1996}), \bibinfo{pages}{123--140}.
\newblock


\bibitem[\protect\citeauthoryear{Chawla, Bowyer, Hall, and Kegelmeyer}{Chawla
  et~al\mbox{.}}{2002}]%
        {SMOTE}
\bibfield{author}{\bibinfo{person}{N.V. Chawla}, \bibinfo{person}{K.W. Bowyer},
  \bibinfo{person}{L.O. Hall}, {and} \bibinfo{person}{W.P. Kegelmeyer}.}
  \bibinfo{year}{2002}\natexlab{}.
\newblock \showarticletitle{SMOTE: Synthetic Minority Over-Sampling Technique}.
\newblock \bibinfo{journal}{\emph{Journal of Artificial Intelligence Research}}
  (\bibinfo{year}{2002}).
\newblock


\bibitem[\protect\citeauthoryear{Gholami, Ford, Fang, Plumptre, Tambe, Driciru,
  Wanyama, Rwetsiba, Nsubaga, and Mabonga}{Gholami et~al\mbox{.}}{2017}]%
        {gholami2017taking}
\bibfield{author}{\bibinfo{person}{Shahrzad Gholami}, \bibinfo{person}{Benjamin
  Ford}, \bibinfo{person}{Fei Fang}, \bibinfo{person}{Andrew Plumptre},
  \bibinfo{person}{Milind Tambe}, \bibinfo{person}{Margaret Driciru},
  \bibinfo{person}{Fred Wanyama}, \bibinfo{person}{Aggrey Rwetsiba},
  \bibinfo{person}{Mustapha Nsubaga}, {and} \bibinfo{person}{Joshua Mabonga}.}
  \bibinfo{year}{2017}\natexlab{}.
\newblock \showarticletitle{Taking it for a test drive: a hybrid
  spatio-temporal model for wildlife poaching prediction evaluated through a
  controlled field test}. In \bibinfo{booktitle}{\emph{Joint European
  Conference on Machine Learning and Knowledge Discovery in Databases}}.
  Springer, \bibinfo{pages}{292--304}.
\newblock


\bibitem[\protect\citeauthoryear{Gholami, Mc~Carthy, Dilkina, Plumptre, Tambe,
  Driciru, Wanyama, Rwetsiba, Nsubaga, Mabonga, et~al\mbox{.}}{Gholami
  et~al\mbox{.}}{2018}]%
        {gholami2018adversary}
\bibfield{author}{\bibinfo{person}{Shahrzad Gholami}, \bibinfo{person}{Sara
  Mc~Carthy}, \bibinfo{person}{Bistra Dilkina}, \bibinfo{person}{Andrew
  Plumptre}, \bibinfo{person}{Milind Tambe}, \bibinfo{person}{Margaret
  Driciru}, \bibinfo{person}{Fred Wanyama}, \bibinfo{person}{Aggrey Rwetsiba},
  \bibinfo{person}{Mustapha Nsubaga}, \bibinfo{person}{Joshua Mabonga},
  {et~al\mbox{.}}} \bibinfo{year}{2018}\natexlab{}.
\newblock \showarticletitle{Adversary models account for imperfect crime data:
  Forecasting and planning against real-world poachers}.
\newblock  (\bibinfo{year}{2018}).
\newblock


\bibitem[\protect\citeauthoryear{Kang and Kang}{Kang and Kang}{2017}]%
        {kang2017prediction}
\bibfield{author}{\bibinfo{person}{Hyeon-Woo Kang} {and}
  \bibinfo{person}{Hang-Bong Kang}.} \bibinfo{year}{2017}\natexlab{}.
\newblock \showarticletitle{Prediction of crime occurrence from multi-modal
  data using deep learning}.
\newblock \bibinfo{journal}{\emph{PloS one}} \bibinfo{volume}{12},
  \bibinfo{number}{4} (\bibinfo{year}{2017}), \bibinfo{pages}{e0176244}.
\newblock


\bibitem[\protect\citeauthoryear{Kar, Ford, Gholami, Fang, Plumptre, Tambe,
  Driciru, Wanyama, Rwetsiba, Nsubaga, et~al\mbox{.}}{Kar
  et~al\mbox{.}}{2017}]%
        {kar2017cloudy}
\bibfield{author}{\bibinfo{person}{Debarun Kar}, \bibinfo{person}{Benjamin
  Ford}, \bibinfo{person}{Shahrzad Gholami}, \bibinfo{person}{Fei Fang},
  \bibinfo{person}{Andrew Plumptre}, \bibinfo{person}{Milind Tambe},
  \bibinfo{person}{Margaret Driciru}, \bibinfo{person}{Fred Wanyama},
  \bibinfo{person}{Aggrey Rwetsiba}, \bibinfo{person}{Mustapha Nsubaga},
  {et~al\mbox{.}}} \bibinfo{year}{2017}\natexlab{}.
\newblock \showarticletitle{Cloudy with a chance of poaching: adversary
  behavior modeling and forecasting with real-world poaching data}. In
  \bibinfo{booktitle}{\emph{Proceedings of the 16th Conference on Autonomous
  Agents and MultiAgent Systems}}. International Foundation for Autonomous
  Agents and Multiagent Systems, \bibinfo{pages}{159--167}.
\newblock


\bibitem[\protect\citeauthoryear{Kim and Ghosh}{Kim and Ghosh}{2017}]%
        {Taewan}
\bibfield{author}{\bibinfo{person}{Taewan Kim} {and} \bibinfo{person}{Joydeep
  Ghosh}.} \bibinfo{year}{2017}\natexlab{}.
\newblock \showarticletitle{Relaxed Oracles for Semi-Supervised Clustering}.
\newblock \bibinfo{journal}{\emph{arXiv preprint arXiv:1711.07433}}
  (\bibinfo{year}{2017}).
\newblock


\bibitem[\protect\citeauthoryear{LeCun, Bengio, and Hinton}{LeCun
  et~al\mbox{.}}{2015}]%
        {lecun2015deep}
\bibfield{author}{\bibinfo{person}{Yann LeCun}, \bibinfo{person}{Yoshua
  Bengio}, {and} \bibinfo{person}{Geoffrey Hinton}.}
  \bibinfo{year}{2015}\natexlab{}.
\newblock \showarticletitle{Deep learning}.
\newblock \bibinfo{journal}{\emph{nature}} \bibinfo{volume}{521},
  \bibinfo{number}{7553} (\bibinfo{year}{2015}), \bibinfo{pages}{436}.
\newblock


\bibitem[\protect\citeauthoryear{Lemieux}{Lemieux}{2014}]%
        {lemieux2014situational}
\bibfield{author}{\bibinfo{person}{Andrew~M Lemieux}.}
  \bibinfo{year}{2014}\natexlab{}.
\newblock \bibinfo{booktitle}{\emph{Situational prevention of poaching}}.
  Vol.~\bibinfo{volume}{15}.
\newblock \bibinfo{publisher}{Routledge}.
\newblock


\bibitem[\protect\citeauthoryear{Moore, Mulindahabi, Masozera, Nichols, Hines,
  Turikunkiko, and Oli}{Moore et~al\mbox{.}}{2017}]%
        {moore2017ranger}
\bibfield{author}{\bibinfo{person}{Jennifer~F Moore}, \bibinfo{person}{Felix
  Mulindahabi}, \bibinfo{person}{Michel~K Masozera}, \bibinfo{person}{James~D
  Nichols}, \bibinfo{person}{James~E Hines}, \bibinfo{person}{Ezechiel
  Turikunkiko}, {and} \bibinfo{person}{Madan~K Oli}.}
  \bibinfo{year}{2017}\natexlab{}.
\newblock \showarticletitle{Are ranger patrols effective in reducing
  poaching-related threats within protected areas?}
\newblock \bibinfo{journal}{\emph{Journal of Applied Ecology}}
  (\bibinfo{year}{2017}).
\newblock


\bibitem[\protect\citeauthoryear{Nguyen, Sinha, Gholami, Plumptre, Joppa,
  Tambe, Driciru, Wanyama, Rwetsiba, Critchlow, et~al\mbox{.}}{Nguyen
  et~al\mbox{.}}{2016}]%
        {nguyen2016capture}
\bibfield{author}{\bibinfo{person}{Thanh~H Nguyen}, \bibinfo{person}{Arunesh
  Sinha}, \bibinfo{person}{Shahrzad Gholami}, \bibinfo{person}{Andrew
  Plumptre}, \bibinfo{person}{Lucas Joppa}, \bibinfo{person}{Milind Tambe},
  \bibinfo{person}{Margaret Driciru}, \bibinfo{person}{Fred Wanyama},
  \bibinfo{person}{Aggrey Rwetsiba}, \bibinfo{person}{Rob Critchlow},
  {et~al\mbox{.}}} \bibinfo{year}{2016}\natexlab{}.
\newblock \showarticletitle{Capture: A new predictive anti-poaching tool for
  wildlife protection}. In \bibinfo{booktitle}{\emph{Proceedings of the 2016
  International Conference on Autonomous Agents \& Multiagent Systems}}.
  International Foundation for Autonomous Agents and Multiagent Systems,
  \bibinfo{pages}{767--775}.
\newblock


\bibitem[\protect\citeauthoryear{Nitesh V.~Chawla}{Nitesh V.~Chawla}{2005}]%
        {Nitesh}
\bibfield{author}{\bibinfo{person}{Grigoris~Karakoulas Nitesh V.~Chawla}.}
  \bibinfo{year}{2005}\natexlab{}.
\newblock \showarticletitle{Learning From Labeled And Unlabeled Data: An
  Empirical Study Across Techniques And Domains}.
\newblock \bibinfo{journal}{\emph{Journal of Artificial Intelligence Research}}
  (\bibinfo{year}{2005}).
\newblock
\showISSN{331-366}


\bibitem[\protect\citeauthoryear{Saif and MacMillan}{Saif and
  MacMillan}{2016}]%
        {saif2016poaching}
\bibfield{author}{\bibinfo{person}{Samia Saif} {and}
  \bibinfo{person}{Douglas~Craig MacMillan}.} \bibinfo{year}{2016}\natexlab{}.
\newblock \showarticletitle{Poaching, trade, and consumption of tiger parts in
  the Bangladesh Sundarbans}.
\newblock In \bibinfo{booktitle}{\emph{The Geography of Environmental Crime}}.
  \bibinfo{publisher}{Springer}, \bibinfo{pages}{13--32}.
\newblock


\bibitem[\protect\citeauthoryear{Scriber}{Scriber}{2014}]%
        {NGpoaching2014}
\bibfield{author}{\bibinfo{person}{Brad Scriber}.}
  \bibinfo{year}{2014}\natexlab{}.
\newblock \bibinfo{title}{100,000 Elephants Killed by Poachers in Just Three
  Years, Landmark Analysis Finds}.
\newblock
\newblock
\newblock
\shownote{\url{https://news.nationalgeographic.com/news/2014/08/140818-elephants-africa-poaching-cites-census/}.}


\bibitem[\protect\citeauthoryear{Shaffer and Bishop}{Shaffer and
  Bishop}{2016}]%
        {shaffer2016predicting}
\bibfield{author}{\bibinfo{person}{Michael~J Shaffer} {and}
  \bibinfo{person}{Joseph~A Bishop}.} \bibinfo{year}{2016}\natexlab{}.
\newblock \showarticletitle{Predicting and preventing elephant poaching
  incidents through statistical analysis, gis-based risk analysis, and aerial
  surveillance flight path modeling}.
\newblock \bibinfo{journal}{\emph{Tropical Conservation Science}}
  \bibinfo{volume}{9}, \bibinfo{number}{1} (\bibinfo{year}{2016}),
  \bibinfo{pages}{525--548}.
\newblock


\bibitem[\protect\citeauthoryear{Shojaee, Mustapha, Sidi, and Jabar}{Shojaee
  et~al\mbox{.}}{2013}]%
        {shojaee2013study}
\bibfield{author}{\bibinfo{person}{Somayeh Shojaee}, \bibinfo{person}{Aida
  Mustapha}, \bibinfo{person}{Fatimah Sidi}, {and} \bibinfo{person}{Marzanah~A
  Jabar}.} \bibinfo{year}{2013}\natexlab{}.
\newblock \showarticletitle{A study on classification learning algorithms to
  predict crime status}.
\newblock \bibinfo{journal}{\emph{International Journal of Digital Content
  Technology and its Applications}} \bibinfo{volume}{7}, \bibinfo{number}{9}
  (\bibinfo{year}{2013}), \bibinfo{pages}{361}.
\newblock


\bibitem[\protect\citeauthoryear{Speirs‐Bridge, Fidler, McBride, Flander,
  Cumming, and Burgman}{Speirs‐Bridge et~al\mbox{.}}{2010}]%
        {4point}
\bibfield{author}{\bibinfo{person}{Andrew Speirs‐Bridge},
  \bibinfo{person}{Fiona Fidler}, \bibinfo{person}{Marissa McBride},
  \bibinfo{person}{Louisa Flander}, \bibinfo{person}{Geoff Cumming}, {and}
  \bibinfo{person}{Mark Burgman}.} \bibinfo{year}{2010}\natexlab{}.
\newblock \showarticletitle{SMOTE: Synthetic Minority Over-Sampling Technique}.
\newblock \bibinfo{journal}{\emph{Risk Analysis}}  \bibinfo{volume}{16}
  (\bibinfo{year}{2010}), \bibinfo{pages}{2098}.
\newblock


\bibitem[\protect\citeauthoryear{VOA}{VOA}{2017}]%
        {VOA2017}
\bibfield{author}{\bibinfo{person}{VOA}.} \bibinfo{year}{2017}\natexlab{}.
\newblock \bibinfo{title}{Poachers Target Africa's Lions, Vultures With
  Poison}.
\newblock
\newblock
\newblock
\shownote{\url{https://www.voanews.com/a/poachers-target-african-lions-vultures-with-poison/4046307.html}.}


\bibitem[\protect\citeauthoryear{Yu}{Yu}{2007}]%
        {TingYu}
\bibfield{author}{\bibinfo{person}{Ting Yu}.} \bibinfo{year}{2007}\natexlab{}.
\newblock \emph{\bibinfo{title}{Incorporating Prior Domain Knowledge into
  Inductive Machine Learning Its implementation in contemporary capital
  markets}}.
\newblock \bibinfo{thesistype}{Ph.D. Dissertation}. \bibinfo{school}{University
  of Technology Sydney}, \bibinfo{address}{Sydney, Australia}.
\newblock


\bibitem[\protect\citeauthoryear{Zhipeng~Luo}{Zhipeng~Luo}{2017}]%
        {ZhipengLuo}
\bibfield{author}{\bibinfo{person}{Milos~Hauskrecht Zhipeng~Luo}.}
  \bibinfo{year}{2017}\natexlab{}.
\newblock \showarticletitle{Active Learning of Classification Models from
  Soft-Labeled Groups}. LLD Workshop 2017.
\newblock


\end{thebibliography}
\newpage
\section{Appendix} \label{appendix}

In the Evaluation section, we discussed about the performance of the models on the data collected between 2013-2018. This includes the data collected during the field test in 2018. This field test was conducted using predictions of poaching activities made by \emph{DT with DD and NS} trained on 2013-2017 dataset. In this section we show results of the models trained on 2013-2017 dataset to illustrate the performance of our models on data collected purely by human judgment. The results of these models are shown in Table \ref{tab:17scores}. We observe that the results on this dataset is better on average for all models.
\begin{table}
\caption {Scores for our models on 2013-2017 dataset} \label{tab:17scores}
\begin{center}
\begin{tabular}{ |c|c|c|c|c|} 
 \hline
Models & LL score & Recall & Precision & F1 score\\
\hline
DT with DD & 17.76 & 0.30 & 0.23 & 0.260\\ 
DT with SMOTE & 17.01 & 0.36 & 0.16& 0.222\\
DT with DD, PS & 18.8 & 0.32 & 0.22 & 0.261\\   
DT with DD and NS & 18.4 & 0.27 & 0.26 & 0.265\\ 
DT with DD, NS, PS & 19.62 & 0.29 & 0.25 & 0.269\\
NN with DD & 3.15 & 0.7 & 0.017 & 0.033\\
NN with DD, NS & 2.14 & 0.44 & 0.018 & 0.035\\
NN with DD, NS, PS & 3.53 & 0.73 & 0.019 & 037 \\
\hline
\end{tabular}
\end{center}
\end{table}
\end{document}